\documentclass[ba,preprint]{imsart}% use this for supplement article
\pubyear{2022}
\volume{TBA}
\issue{TBA}
\arxiv{2105.11022}
\firstpage{1}
\lastpage{1}

\usepackage{amsthm}
\usepackage{amsmath}
\usepackage{natbib}
\usepackage[colorlinks,citecolor=blue,urlcolor=blue,filecolor=blue,backref=page]{hyperref}
\usepackage{graphicx}

\startlocaldefs
\numberwithin{equation}{section}
\theoremstyle{plain}

\endlocaldefs

\usepackage[british]{babel}

\usepackage{enumitem}
\usepackage{graphicx}
\usepackage{amsmath}
\usepackage{amssymb}
\usepackage{bm,lscape}
\usepackage{bbm}
\usepackage{mathrsfs,dsfont}
\usepackage{wasysym}
\usepackage{graphicx}
\usepackage{hyperref}
\usepackage{algorithm}
\usepackage{algorithmic}
\usepackage{xcolor}
\usepackage{adjustbox}

\newcommand{\bbeta}{\boldsymbol{\beta}}

\begin{document}

\begin{frontmatter}

\title{Robust Bayesian Nonparametric Variable Selection for Linear Regression
% \support{Support information of the article.}
}
%\runtitle{Robust Bayesian Nonparametric Variable Selection for Linear Regression}

\begin{aug}
\author{\fnms{Alberto} \snm{Cabezas}\thanksref{addr1}\ead[label=e1]{a.cabezasgonzalez@lancaster.ac.uk}},
\author{\fnms{Marco} \snm{Battiston}\thanksref{addr1}\ead[label=e2]{m.battiston@lancaster.ac.uk}}
\and
\author{\fnms{Christopher} \snm{Nemeth}\thanksref{addr1}%
\ead[label=e3]{c.nemeth@lancaster.ac.uk}}

%\runauthor{A. Cabezas et al.}

\address[addr1]{Department of Mathematics and Statistics, Lancaster University
    \printead{e1} % print email address of "e1"
    % \printead*{e2}
}

% \address[addr2]{Address of the Third author
%     Usually a few lines long
%     Usually a few lines long
%     \printead{e3}
%     \printead{u1}
% }

% \thankstext{t1}{Some comment}
% \thankstext{t2}{First supporter of the project}
% \thankstext{t3}{Second supporter of the project}

\end{aug}

\begin{abstract}
  Spike-and-slab and horseshoe regression are arguably the most popular Bayesian variable selection approaches for linear regression models. However, their performance can deteriorate if outliers and heteroskedasticity are present in the data, which are common features in many real-world statistics and machine learning applications. In this work, we propose a Bayesian nonparametric approach to linear regression that performs variable selection while accounting for outliers and heteroskedasticity. Our proposed model is an instance of a Dirichlet process scale mixture model with the advantage that we can derive the full conditional distributions of all parameters in closed form, hence producing an efficient Gibbs sampler for posterior inference. Moreover, we present how to extend the model to account for heavy-tailed response variables. The performance of the model is tested against competing algorithms on synthetic and real-world datasets.
\end{abstract}

%% ** Keywords **

% \begin{keyword}[class=MSC]
% \kwd[Primary ]{60K35}
% \kwd{60K35}
% \kwd[; secondary ]{60K35}
% \end{keyword}

\end{frontmatter}

\section{Introduction}

%Spike-and-slab and horseshoe regression are popular algorithms for Bayesian variable selection. These algorithms address the challenge of fitting a large number of attributes to response data by identifying a restricted number of key attributes. 

Bayesian variable selection is a popular tool in statistics and machine learning that can be used for feature selection in linear regression models. 
The two most popular models are arguably the spike-and-slab and horseshoe regression models. Both models rely on choosing a sparsity inducing prior over the regression coefficients $\pmb{\beta} \in \mathbb{R}^p$. In the spike-and-slab model, the prior is a mixture between a point mass at zero and a diffuse prior, while in the horseshoe model, a continuous prior balances the local and global shrinkage (see Section \ref{sec2}). Compared to the estimator for the regression coefficients in standard linear regression, the posterior distribution under these priors produces a shrinkage effect toward zero in the posterior point estimates. This is especially useful in the $p>>n$ regime, in which the number of attributes $p$ can be much larger than the number of observations $n$.
% In the case of spike-and-slab regression, a prior distribution over the regression coefficients $\pmb{\beta} \in \mathbb{R}^p$ is given as a mixture between a point mass at zero, \emph{spike}, \textit{(which turns off the associated attribute)} and a diffuse prior, \emph{slab} \textit{(which turns on the associated attribute)}. The posterior distribution conditional on the data will also display a similar mixture structure, hence producing a shrinkage effect toward zero in the posterior point estimates of the regression coefficients compared to the standard estimator in linear regression. This is especially usefuamp;ampels usually assume the same conditional variance $\sigma^{2}$ for each observation. This assumption can dramatically deteriorate the posterior estimates of the regression coefficients in the presence of heteroskedasticity, particularly when the number of attributes is large. 

In this paper we propose extensions of the spike-and-slab and horseshoe regression models to deal with heteroskedasticity and outliers in the data. Specifically, we propose a Bayesian nonparametric model in which each observation is endowed with its own specific variance $\sigma_{i}^{2}$, which is sampled from an unknown distribution $P$. The distribution $P$ is a Dirichlet process prior and the resulting model is a Dirichlet process scale mixture model. Due to the discreteness of the Dirichlet process, the resulting vector of variances $(\sigma^{2}_{1},\ldots,\sigma^{2}_{n})$ will be partitioned into groups, hence also producing a corresponding clustering of the observations, in which observations belonging to the same group will have the same conditional variance. Moreover, some observations may be allocated to a single group having much larger variance than the others, hence allowing for outliers in the data.

The key features of our proposed model are:
\begin{itemize}[noitemsep]
\item \textbf{Parsimonious regression construction:}
Our nonparametric method performs
feature selection with the posterior concentrating on the most relevant
prediction coefficients. However, unlike variable coefficient models,
there is no increase in the degrees-of-freedom associated with the
number of regression parameters (or smoothness of parameters to control,
c.f. functional regression).
\item \textbf{Interpretable model structure:} The proposed linear regression model changes the structure of the marginal variance components while retaining the highly interpretable properties of a standard sparse regression formulation.
\item \textbf{Posterior credible intervals:} A fully Bayesian approach gives credible intervals for the regression coefficients. We propose several ways to perform posterior inference selection using these intervals, which provide uncertainty quantification for decision makers working with high-dimensional data.
\item \textbf{Efficient inference:} Under our model, full conditional distributions of all parameters can be derived in \emph{closed form}, hence producing an efficient Gibbs algorithm that gives a tuning-free and rejection-free Markov chain Monte Carlo (MCMC) sampler.
\end{itemize}

The rest of the paper is organized as follows. Section \ref{sec2} briefly reviews the spike-and-slab and horseshoe regression models. Section \ref{sec:robustSVS} introduces the Bayesian nonparametric framework and our Dirichlet process mixture model for linear regression along with details of our MCMC sampler. In Subsection \ref{sec:t-model}, we provide an extension to our model to account for heavy-tailed response variables.
Finally, Section \ref{sec4} presents an empirical comparison of our proposed model against popular alternative models on synthetic data and real-world data examples. 

\section{Bayesian Variable Selection for Linear Regression} \label{sec2}

The Bayesian approach to variable selection for linear regression models is to introduce sparsity-inducing priors on the regression coefficients. The two most popular approaches in the literature, which we shall review here, are the discrete mixture priors known as the spike-and-slab \citep{mitchell1988bayesian, george1993variable}, and the continuous shrinkage priors, most notably the horseshoe prior \citep{carvalho2010horseshoe}. 

\subsection{Spike-and-slab priors} \label{sec2.1}
The original spike-and-slab model was initially proposed by Mitchell \& Beauchamp (1988) and significantly developed by Madigan \& Raftery (1994) and George \& McCulloch (1997). The final adjustments to the model were completed by Ishwaran \& Rao (2005) in what they refer to as the \textit{stochastic variable selection} model. The spike-and-slab prior is intuitively simple and consists of two components. The \textit{spike} is a delta function centered at zero indicating $\beta_j \approx 0$, and the \textit{slab} gives probability mass to non-zero coefficients.  

In our regression framework, we have data $\textbf{y}\in \mathbb{R}^n$ that is explained by a matrix of attributes $\textbf{X} \in \mathbb{R}^{n \times p}$ and coefficients $\boldsymbol{\beta} \in \mathbb{R}^p$. Assuming a spike-and-slab prior for $\boldsymbol{\beta}$, we have the following model,
\begin{align} \label{SVS}
y_{i}|\boldsymbol{x}_{i},\boldsymbol{\beta},\sigma^{2}\sim \mathcal{N}(\boldsymbol{x}_{i}^{\top}\boldsymbol{\beta},\sigma^{2}) &  & i=1,\ldots,n\nonumber \\
\beta_{j}|\eta_{j},\tau_{j}^{2}\sim\mathcal{N}(0,\eta_{j}\tau_{j}^{2}) &  & j=1,\ldots,p\nonumber \\
\eta_{j}|\nu_{0},\omega \sim (1-\omega)\delta_{\nu_{0}}(\cdot)+\omega\delta_{1}(\cdot)\nonumber \\
\tau_{j}^{2}|a_{1},a_{2}\sim \mathrm{IG}(a_{1},a_{2})\\
\omega\sim\mathrm{Unif}[0,1]\nonumber \\
\sigma^{2}|b_{1},b_{2}\sim \mathrm{IG}(b_{1},b_{2})\nonumber 
\end{align}

The likelihood for the data assumes a standard Gaussian regression model, and the prior for $\boldsymbol{\beta}$ is a scale
mixture of Gaussians. The variable $\eta_{j}$ is a latent indicator and $\delta_{v_{0}}(\cdot)$ denotes a point mass at $v_{0}$, where $v_{0}$ is a value chosen close to $0$, and thus the variance of the
prior on $\boldsymbol{\beta}$ is either %a concentrated ($v_{0}\tau_{j}^{2}$)
almost zero or a broad % ($\tau_{j}^{2}$) 
Gaussian distribution. 

\subsection{Horseshoe priors} 
\label{sec2.2}
Spike-and-slab priors are intuitively appealing, but in practice their discrete nature makes posterior inference computationally difficult. A popular alternative perspective on sparse Bayesian regression is given by the horseshoe prior construction \citep[see]{carvalho2009handling,carvalho2010horseshoe}. The horseshoe prior is a continuous shrinkage prior which makes posterior computation more efficient when using gradient-based MCMC sampling tools such as STAN \citep{carpenter2017stan}.
\vspace{-0.1cm}
\begin{align} \label{horseshoe_}
y_{i}|\boldsymbol{x}_{i},\boldsymbol{\beta},\sigma^{2}\sim \mathcal{N}(\boldsymbol{x}_{i}^{\top}\boldsymbol{\beta},\sigma^{2}) &  & i=1,\ldots,n\nonumber \\
\beta_{j}|\lambda_{j}^2,\tau^{2} \sim \mathcal{N}(0,\lambda_{j}^2\tau^{2}) &  & j=1,\ldots,p\nonumber \\
\lambda_{j} \sim \mathcal{C}^{+}(0,1)  \nonumber \\
\tau \sim \mathcal{C}^{+}(0,1) \\
\sigma^{2}|b_{1},b_{2}\sim \mathrm{IG}(b_{1},b_{2}) \nonumber 
\end{align}

Under the horseshoe regression model, the parameters $\lambda_j$ and $\tau$ are the \textit{local} and \textit{global} shrinkage parameters, respectively. Following \cite{carvalho2010horseshoe}, we choose half-Cauchy priors for $\lambda_j$ and $\tau$. 
The intuition behind the horseshoe prior is that the global parameter $\tau$ will force the regression coefficients towards zero, while the heavy tails of the half-Cauchy prior for the local shrinkage parameters $\lambda_j$ will allow for non-zero $\bbeta$ coefficients. 

The horseshoe model used in our experiments follows the form originally proposed by \cite{carvalho2010horseshoe}. A standard Gibbs sampling approach on this parametrization of the model is difficult to implement due to the non-conjugate posterior form of the shrinkage parameters in the model. In our implementation, we introduce auxiliary variables that lead to conjugate full conditionals for all parameters as suggested by \cite{makalic2015}, and allow for a straightforward implementation of Gibbs sampling.

The parameterization given by \cite{makalic2015} makes use of an inverse gamma scale mixture representation of a random variable with a half-Cauchy distribution. It can be shown that $X \sim \mathcal{C}^{+}(0,A)$ if $X^2|a \sim \mathrm{IG}(1/2,1/a)$ and $a \sim \mathrm{IG}(1/2,1/A^2)$. Using this decomposition leads to the reparametrized horseshoe model
\begin{align} \label{horseshoe}
y_{i}|\boldsymbol{x}_{i},\boldsymbol{\beta},\sigma^{2}\sim \mathcal{N}(\boldsymbol{x}_{i}^{\top}\boldsymbol{\beta},\sigma^{2}) &  & i=1,\ldots,n\nonumber \\
\beta_{j}|\lambda_{j}^2,\tau^{2} \sim \mathcal{N}(0,\lambda_{j}^2\tau^{2}) &  & j=1,\ldots,p\nonumber \\
\lambda_{j}^2 \sim \mathrm{IG}(1/2,1/\nu_j)  \nonumber \\
\tau^2 \sim \mathrm{IG}(1/2,1/\xi) \\
\nu_1,\dots,\nu_p,\xi \sim \mathrm{IG}(1/2,1) \nonumber \\
\sigma^{2}|b_{1},b_{2}\sim \mathrm{IG}(b_{1},b_{2}), \nonumber 
\end{align}
for which the sampling schemes presented in the next section easily follow. 

\section{Nonparametric Stochastic Variable Selection} \label{sec:robustSVS}

\subsection{Background to Dirichlet processes} 
\label{sec3.1}

In Bayesian nonparametric statistics, unknown parameters are infinite dimensional and cannot be parametrized by a subset of Euclidean space. The most common examples of nonparametric problems are the estimation of density functions, distribution functions and nonparametric regression \citep{hjo10}.

The most popular nonparametric distribution function is the Dirichlet process, introduced in \cite{Fer73}. In this setting, $P$ has a \emph{Dirichlet process} distribution with base measure $P_{0}$ and concentration parameter $\alpha$, denoted $P\sim DP(\alpha,P_{0})$. For every measurable partition $(A_{1},\ldots,A_{k})$, the random vector $(P(A_{1}),\dots,P(A_{k}))$ has a Dirichlet distribution on the $k$-dimensional simplex with parameters $(\alpha P_{0}(A_{1}),\ldots,\alpha P_{0}(A_{k}))$. 
The base measure $P_{0}$ is the mean of the prior (i.e. $\mathbb{E}(P)=P_{0}$), while the concentration parameter $\alpha$ regulates prior uncertainty around $P_{0}$. A large $\alpha$ value implies a strong belief in the prior.
%A random measure $P$ on the sample space $\mathcal{X}$ possesses a {\normalfont Dirichlet process} distribution with base measure $P_{0}$ and concentration parameter $\alpha$, denoted $P\sim DP(\alpha,P_{0})$, if, for every measurable partition $(A_{1},\ldots,A_{k})$ of $\mathcal{X}$, the random vector $(P(A_{1}),\ldots,P(A_{k}))$ has a Dirichlet distribution on the $k$ dimensional simplex with parameters $(\alpha P_{0}(A_{1}),\ldots,\alpha P_{0}(A_{k}))$.

There are many different representations of the Dirichlet process (see \cite{Gho18}, Chapter 4). For the purpose of this paper, we shall utilize the \textit{Chinese restaurant process representation}, which gives the marginal distribution of the data when the unknown distribution $P$ has been marginalized out. %Specifically, if we consider the following exchangeable model for a generic variable $\theta$,
%\begin{align*}
%    \theta_{i}|P & \stackrel{iid}{\sim} P \ \ i=1,\ldots,n \\
%    P &\sim DP(\alpha,P_{0}),
%\end{align*}
Specifically, if $\theta_{1:n}$ is an i.i.d. sample from $P$, $\theta_{i}|P  \stackrel{iid}{\sim} P$, and $P \sim DP(\alpha,P_{0})$, then
the marginal distribution of $\theta_{1:n}$, when $P$ is integrated out, is
\begin{align*}
  \pi(\theta_{1:n}|\alpha,P_{0})\nonumber  =\int\prod_{i=1}^{n}\mathcal{P}(\theta_{i})\mathrm{DP}(d\mathcal{P};\alpha,P_{0}),
\end{align*}
which can be described by the marginal of $\theta_{1}$ and the conditional distributions of $\theta_{i}|\theta_{1:i-1}$ for $i=2,\ldots,n$,
 $\pi(\theta_{1:n}|\alpha,P_{0}) = \pi(\theta_{1}|\alpha,P_{0})\prod_{i=2}^{n}\pi(\theta_{i}|\theta_{1:i-1},\alpha,P_{0})$,
where $\pi(\theta_{1}|\alpha,P_{0})=P_{0}$ and
\begin{align*}
\theta_{i}|\theta_{1:i-1},\alpha,P_{0} \sim  \frac{1}{i-1+\alpha}\sum_{i=1}^{n}\delta_{\theta_{i}} +\frac{\alpha}{i-1+\alpha}P_{0}.
\end{align*}
Thus, a sample $\theta_{1:n}$ from $\pi(\theta_{1:n}|\alpha,P_{0})$
will display ties with positive probability. The parameter $\alpha$ regulates the number of clusters in  $\theta_{1:n}$. All else being equal,
larger $\alpha$ will lead to more distinct values within $\theta_{1:n}$.
%%%

\subsection{Dirichlet process mixture model}
\label{sec:dppmix}

The Dirichlet process mixture model (DPM) \citep{Lo84} assumes each observation $y_{i}$ is sampled from a countable mixture model. The likelihood of each mixture component, $K(y_{i};\theta^{*}_{k})$, is parametrized by an unknown parameter vector $\theta^{*}_{k}$. In the following, $K$ will be a Gaussian kernel and $\theta^{*}_{k}$ is the group specific variance. Both the mixture parameters $\theta^{*}$ and the mixture weights $w_{k}$ can be encoded into a random measure $P=\sum_{k\geq 1}w_{k}\delta_{\theta^{*}_{k}}$ which is endowed with a Dirichlet process prior, hence producing the following mixture model for $i=1,\ldots,n$ 
\begin{align*}
    y_{i}|P &\sim \sum_{k=1}^{\infty} w_{k} K(y_{i};\theta^{*}_{k}) = \int K(y_{i};\theta) P(d\theta) \\
     P &\sim DP(\alpha,P_{0}).
\end{align*}
By introducing latent class variables, $\{\theta_{i}\}_{i=1}^{n}$ s.t. $\mathbb{P}(\theta_{i}=\theta^{*}_{k})=w_{k}$, the DPM model is equivalent to the latent variable mixture model
\begin{align} \label{DPmixt}
    y_{i}|\theta_{i} &\sim  K(y_{i};\theta_{i}) \qquad & i=1,\ldots,n \nonumber \\ 
    \theta_{i}|P &\overset{iid}{\sim }P \qquad \ \ \ & i=1,\ldots,n \\
    P &\sim DP(\alpha,P_{0}) & \nonumber
\end{align}
The DPM model has been used in a variety of applications in statistics and machine learning for density estimation and clustering. In density estimation, the goal is to estimate the unknown density of the observations through a mixture model, while in clustering the goal is to cluster observations into groups with a similar distribution. For the latter task, the discreteness property of the Dirichlet process is very convenient, since with positive probability, we will observe ties among the latent variables $\theta_{1:n}$. Two observations $y_{i}$ and $y_{j}$ having the same value of the latent variables $\theta_{i}$ and $\theta_{j}$ will be assigned to the same cluster and have the same distribution. %There are many computational algorithms to perform inference in DPM models, including MCMC, variational inference and Newton recursion algorithms (see  Chapter 5 of \cite{Gho18}).

In Section \ref{sec:dp-variable} we utilize the properties of the Dirichlet process mixture model to propose a Bayesian nonparametric extension to the spike-and-slab and horseshoe regression models. We show that our new model is able to capture heteroskedasticity and outliers in the observations, as well as capturing clustering behaviour in the variance structure. Under both DP variable selection models the mixture component has a likelihood with a cluster-specific variance term which is not fixed \textit{apriori}, but is learned from the data. 

%Our proposed methodology builds on the spike-and-slab SVS model by \cite{Ishwaran2005} described in Subsection \ref{sec2.1}. In order to accommodate heteroskedasity and outliers, we propose a nonparametric extension using a DP scale mixture model to capture clustering behaviour in the variance structure. In the proposed \emph{Heteroskedastic Stochastic Variable Selection} (HSVS) model each mixture component has a SVS likelihood, but with cluster specific variance. The number of variance clusters is not fixed a priori, but learnt from the data.

\subsection{Dirichlet process variable selection}
\label{sec:dp-variable}

In both the spike-and-slab \eqref{SVS} and horsehoe \eqref{horseshoe} models, one assumes that each $y_{i}$ has the same conditional variance $\sigma^{2}$. Under our nonparametric Dirichlet process model, we introduce an observation dependent variance $\sigma_{i}^{2}$ for each data point. The vector $\boldsymbol{\sigma^{2}}:=\sigma_{1:n}^{2}$ is assumed to be sampled from an unknown discrete distribution $P$ sampled from a Dirichlet process. Specifically,
we consider the following general hierarchical model,
\begin{align} \label{HSVS}
y_{i}|\boldsymbol{x}_{i},\boldsymbol{\beta},\boldsymbol{\sigma}^2 & \sim \mathcal{N}(\boldsymbol{x}_{i}^{\top}\boldsymbol{\beta},\sigma_{i}^{2}) &  & i=1,\ldots,n\nonumber \\
\pmb{\beta} & \sim \pi  \\
\sigma_{i}^{2}|P & \sim P(d\sigma_{i}^{2}) &  & i=1,\ldots,n\nonumber \\
P|\alpha & \sim\mathrm{DP}(\alpha,\mathrm{IG}(b_{1},b_{2})) \nonumber \\
\alpha & \sim \text{Gamma}(d_{1},d_{2}), \nonumber
\end{align}

where the prior $\pi$ for $\pmb{\beta}$ either follows the spike-and-slab construction in lines 2-5 of eq. \eqref{SVS}, or the horseshoe model in lines 2-4 of eq. \eqref{horseshoe}. We refer to the general model in eq. \eqref{HSVS} as the \textit{Dirichlet process variable selection model} and recognize that the spike-and-slab and horseshoe models are special cases, which we denote as \textit{Dirichlet process spike-and-slab} (DPSS) and \textit{Dirichlet process horseshoe} (DPHS), respectively.

In this model, we are assuming a variance $\sigma_{i}$ for each observation
$y_{i}$, where the vector of variances $\sigma_{1},\ldots,\sigma_{n}$
is assumed to be conditionally i.i.d. from an unknown distribution
$P$. Specifically, we use a Dirichlet process
as a nonparametric prior for $P$, centered at $\mathrm{IG}(b_{1},b_{2})$ with concentration parameter $\alpha$. From
the properties of the DP, $P$ will almost surely be a discrete
distribution. This implies that, with positive probability, we will observe \emph{ties} among $\sigma_{1},\ldots,\sigma_{n}$. In other words, some $\sigma_i$ will take the same value. We denote by $\sigma_{1}^{*},\ldots,\sigma_{K_{n}}^{*}$ the $K_{n}$ distinct values assumed by $\sigma_{1},\ldots,\sigma_{n}$. We will then have
$K_{n}$ clusters among our observations $\{y_{i}\}_{i=1}^n$, where observations
within a cluster have the same variance, but different clusters have
different variances. Specifically, if $\sigma_{i}=\sigma_{j}$, then
$y_{i}$ and $y_{j}$ will be in the same cluster and have the same
conditional variance, but with possibly different means, depending on their attributes.

\begin{algorithm}
    \caption{DP variable selection Gibbs sampler}
    \label{algo1}
\begin{algorithmic}
  \FOR{$t$ in 1: number of iterations }
  \FOR{i in 1:n }
        \STATE Sample classification value $c_i$ with probability \eqref{c_i}.
        % \STATE Let $K^{-}$ be the number of distinct $c_{j}$ for $j\neq i$,  labeled $\{1,\ldots,K^{-}\}$. 
        
        % Draw new value $c_{i}$ from
        % \begin{align*}
        %     \mathbb{P}[& c_{i} =c\,|\,c_{-i},y_{1:n},\sigma^{2*}_{1:K_{n}}]\propto \\
        %     & \left\{\begin{array}{ll}
        %         n_{-i,c}N(y_{i};\pmb{x}_{i}^{T}\pmb{\beta},\sigma_{c}^{2*})\hspace{0.2cm}\text{for }1\leq c \leq K^{-} \\[0.4cm]
        %         \alpha g(y_{i};\boldsymbol{x}_{i},\boldsymbol{\beta},b_{1},b_{2})\hspace{0.2cm}\text{for  }c=K^{-}+1
        %     \end{array}\right.
        % \end{align*} 
        % where $n_{-i,c}$ is the number of $c_{j}=c$ for $j\neq i$.
        \ENDFOR
   \FOR{k in 1:$K_{n}$}
      \STATE  Sample $\sigma_{k}^{2*}$
       \begin{equation*}
       \sigma_{k}^{2*}\sim\mathrm{IG}\left(b_{1}+\frac{n_{k}}{2},\:b_{2}+\frac{1}{2}\sum_{i:C_{k}}(y_{i}-\boldsymbol{x}_{i}^{\top}\boldsymbol{\beta})^{2}\right),\;\label{eq:variance_sample}
        \end{equation*}
        where $C_k = \{ \sigma^{2}_i \ | \ c_i = k\}$
   \ENDFOR

 \IF{Spike-and-Slab} \STATE Sample $\bbeta$, $\tau_j$, $\eta_j$ and $\omega$ using Algorithm \ref{alg:spike} \ELSIF{Horseshoe} \STATE Sample $\bbeta$, $\lambda_j$ and $\tau$ using Algorithm \ref{alg:horseshoe} \ENDIF

\STATE Sample $\alpha$, by sampling the latent variables
\begin{align*}
        \psi |\alpha,K_{n} &\sim \text{Beta}(\alpha+1,n) \ \ \mbox{then} \\
        a |\psi,K_n &\sim \text{Bernoulli}\left(\frac{w2}{w1+w2}\right) \\
        \alpha |a, \psi,K_{n} &\sim \text{Gamma}(d_{1}+K_{n} + a,d_{2}-\log \psi)
\end{align*}
where $w_1 = d_1 + K_n + 1$ and $w_2= n(d_2-\log\psi)$.
\ENDFOR
\end{algorithmic}
\end{algorithm}

\subsubsection{Posterior inference} 
\label{sec:inference}
Under the DPSS and DPHS models, posterior inference can be carried out efficiently using a Markov chain Monte Carlo sampler. We propose a general Gibbs sampler to sample $\sigma_{1:n}^{2}$ (see Algorithm \ref{algo1}) based on the algorithm by \cite{escobar1995bayesian}, where at each iteration, we first resample a classification vector $c_{1:n}$ assigning each $\sigma_{i}^{2}$ to a block in the partition, and then resample the distinct values $\sigma_{1:k}^{2*}$. 

The partition generated by the variance value assigned to each observation is distributed, \textit{a priori}, as a \textit{Chinese Restaurant Process} \citep{aldous1985exchangeability}. This distribution can be understood intuitively as a Chinese restaurant serving an infinite amount of customers arriving in succession. Each customer needs to be seated on a unbounded number of tables, where the probability of seating a customer at an occupied table is proportional to the number of people already seated at that table. Alternatively, the customer could be seated at a new table with probability proportional to the concentration parameter $\alpha$. Assuming i.i.d. variances, we can assume, \textit{a posteriori}, that each observation is the last customer to arrive and each table represents a cluster of variances. The customer/observation will be seated at an already occupied table, or a new table with probability proportional to %its a priori probability multiplied by the likelihood of that observation having the variance of that table/cluster. Hence, the classification vector $c_{i}$ for each observation $i$ will be drawn with probability
\begin{align}
    \mathbb{P}[& c_{i} =c\,|\,c_{-i},y_{1:n},\sigma^{2*}_{1:K_{n}}]\propto
    \left\{\begin{array}{ll}
        n_{-i,c}N(y_{i};\pmb{x}_{i}^{T}\pmb{\beta},\sigma_{c}^{2*})\hspace{0.2cm}\text{for }1\leq c \leq K^{-} \\  [0.4cm]
        \alpha g(y_{i};\boldsymbol{x}_{i},\boldsymbol{\beta},b_{1},b_{2})\hspace{0.2cm}\text{for  }c=K^{-}+1,
    \end{array}\right. \label{c_i}
\end{align} 
where $K^{-}$ are the number of distinct $c_{j}$ for $j\neq i$, labeled $\{1,\ldots,K^{-}\}$, and $n_{-i,c}$ is the number of $c_{j}=c$ for $j\neq i$. The likelihood of observation $i$ being assigned a new variance, i.e. seated at a new table, is defined as
\begin{align*}
 g(y_{i};\boldsymbol{x}_{i},\boldsymbol{\beta},b_{1},b_{2}) & :=  \int \mathcal{N}(y_{i};\boldsymbol{x}_{i}^{\top}\boldsymbol{\beta},\sigma^{2})\mathrm{IG}(\sigma^{2};b_{1},b_{2})d\sigma^{2}\\
 & =  \frac{b_{2}^{b_{1}}}{\sqrt{2\pi}}\frac{\Gamma(b_{1}+2^{-1})}{\Gamma(b_{1})} \left(\frac{(y_{i}-\boldsymbol{x}_{i}^{\top}\boldsymbol{\beta})^{2}}{2}+b_{2}\right)^{-(b_{1}+\frac{1}{2})}\;.
\end{align*}
Finally, the concentration parameter $\alpha$ is sampled using a data augmentation trick \cite[][see pg.89]{Gho18}.

We sample the regression coefficients $\bbeta$ for the spike-and-slab and horseshoe models using Algorithms \ref{alg:spike} and \ref{alg:horseshoe}, respectively. The Gibbs sampler for the spike-and-slab model follows from \cite{Ishwaran2005} while the Gibbs sampler for the horseshoe model is based on the data augmentation construction of \cite{makalic2015simple}. 
The full conditional distributions of all parameters in both models can be easily derived and have closed form expressions. Alternatively, particularly for the case of $p \gg n$, we can get a random sample from the Multivariate Normal distribution of $\bbeta$ using the linear solver of \citet{bhattacharya2016fast} adapted to our Dirichlet process mixture model. This extension is detailed in Algorithm \ref{alg:fastbeta}, this algorithm would replace the first step of Algorithms \ref{alg:spike} and \ref{alg:horseshoe} once the parameters of matrices $\Sigma$ and $\Lambda$ are fixed. Since both $\Sigma$ and $\Lambda$ are diagonal matrices, the extension changes the cost of sampling $\bbeta$ from $\mathcal{O}(p^3)$ when sampling directly from the multivariate normal to $\mathcal{O}(n^2p)$ when using the linear solver.

An advantage of our Dirichlet process model construction is that the number of clusters $K_n$ that partition the vector of variance parameters $\sigma_{1:n}^{2}$ are learned by the DP model and do not need to be fixed \textit{apriori}. %In Algorithm \ref{algo1}, the conditional probability that $\sigma_{i}^{2}$ is assigned to a new cluster is defined as,

% \begin{align*}
%  g(y_{i};\boldsymbol{x}_{i},\boldsymbol{\beta},b_{1},b_{2}) & :=  \int \mathcal{N}(y_{i};\boldsymbol{x}_{i}^{\top}\boldsymbol{\beta},\sigma^{2})\mathrm{IG}(\sigma^{2};b_{1},b_{2})d\sigma^{2}\\
%  & =  \frac{b_{2}^{b_{1}}}{\sqrt{2\pi}}\frac{\Gamma(b_{1}+2^{-1})}{\Gamma(b_{1})} \cdot \\
%  &  \quad\cdot\left(\frac{(y_{i}-\boldsymbol{x}_{i}^{\top}\boldsymbol{\beta})^{2}}{2}+b_{2}\right)^{-(b_{1}+\frac{1}{2})}\;.
% \end{align*}

%%%% SPIKE AND SLAB SAMPLER
\begin{algorithm}
    \caption{Sample $\beta$ with the spike-and-slab model}
    \label{alg:spike}
\begin{algorithmic}
\STATE   Sample $\boldsymbol{\beta}$
    \begin{equation*}
  \boldsymbol{\beta}\sim \mathcal{N}(\boldsymbol{\mu}_{\beta|\cdot},\Sigma_{\beta|\cdot})
\end{equation*}
where 
$
 \boldsymbol{\mu}_{\beta|\cdot}=(X^{T}\Sigma^{-1}X+\Lambda^{-1})^{-1}X^{T}\Sigma^{-1}\boldsymbol{y}   
$ and
$
  \Sigma_{\beta|\cdot}^{-1}=X^{T}\Sigma^{-1}X+\Lambda^{-1},   
$
with $\Sigma = \mathrm{diag}(\sigma^2_1,\ldots,\sigma^2_p)$ and $\Lambda = \mathrm{diag} (\tau_1^2\eta_1, \ldots, \tau_p^2\eta_p)$,

  \FOR{j in 1:p }
\STATE Sample $\tau_{j}$
$$
    \tau_{j}^{-2}|\bbeta,\eta\sim\mathrm{Gamma}\left(a_{1}+1/2,a_{2}+\beta_{j}^{2}/2\eta_{j}\right),\;
$$
\ENDFOR

\FOR{j in 1:p }
\STATE Sample $\eta_{j}$
$$
    \eta_{j}|\beta,\tau,\omega\sim\frac{w_{1,j}}{w_{1,j}+w_{2,j}}\delta_{v_{0}}(\cdot)+\frac{w_{2,j}}{w_{1,j}+w_{2,j}}\delta_{1}(\cdot)\;,
$$
where $w_{1,j}=(1-\omega)v_{0}^{-1/2}\exp(-\beta_{j}^{2}/2v_{0}\tau_{k}^{2})$
and $w_{2,j}=\omega\exp(-\beta_{j}^{2}/2\tau_{j}^{2})$.
\ENDFOR

\STATE Sample $\omega$
$$
    \omega|\eta,\tau\sim\mathrm{Beta}(1+|\{j\:|\:\eta_{j}=1\}|,\:1+|\{j\:|\:\eta_{j}=v_{0}\}|)
$$

\end{algorithmic}
\end{algorithm}

%%%% HORSESHOE SAMPLER
\begin{algorithm}
    \caption{Sample $\beta$ with the horseshoe model}
    \label{alg:horseshoe}
\begin{algorithmic}
\STATE   Sample $\boldsymbol{\beta}$
    \begin{equation*}
  \boldsymbol{\beta}\sim \mathcal{N}(\boldsymbol{\mu}_{\beta|\cdot},\Sigma_{\beta|\cdot})
\end{equation*}
where 
$
 \boldsymbol{\mu}_{\beta|\cdot}=(X^{T}\Sigma^{-1}X+\Lambda^{-1})^{-1}X^{T}\Sigma^{-1}\boldsymbol{y}$ and
$\Sigma_{\beta|\cdot}^{-1}=X^{T}\Sigma^{-1}X+\Lambda^{-1}$, with $\Sigma = \mathrm{diag}(\sigma^2_1,\ldots,\sigma^2_p)$ and $\Lambda = \mathrm{diag}(\tau^2\lambda^2_1, \ldots, \tau^2\lambda^2_p)$,

  \FOR{j in 1:p }
\STATE Sample the local shrinkage parameter $\lambda^2_{j}$ and auxiliary variable $\nu_j$,

$$
    \lambda^2_{j}| \cdot \sim \mbox{IG}(1,1/\nu_j+\beta_j^2/2\tau^2),
$$

$$
    \nu_j | \cdot \sim \mbox{IG}(1,1+1/\lambda^2_j);
$$

\ENDFOR

\STATE Sample the global shrinkage parameter $\tau^2$ the auxiliary variable $\xi$,
$$
    \tau^2 \sim \mbox{IG}\left((p+1)/2,1/\xi+\sum_{j=1}^p \beta^2_j/2\lambda^2_j\right),
$$
$$
    \xi \sim \mbox{IG}(1,1+1/\tau^2);
$$
\end{algorithmic}
\end{algorithm}

%%%% FAST BETA SAMPLER
\begin{algorithm}
    \caption{Linear solver sampling of $\bbeta$ given $\Sigma$ and $\Lambda$.}
    \label{alg:fastbeta}
\begin{algorithmic}
  \STATE Sample $u \sim \mathcal{N}(0, \Lambda)$ and $\delta \sim \mathcal{N}(0, \mathbb{I}_n)$.
  \STATE Let $v = \Sigma^{-1/2}X + \delta$.
  \STATE Solve $(\Sigma^{-1/2}X\Lambda X^T\Sigma^{-1/2} + \mathbb{I}_n)w = \Sigma^{-1/2}\boldsymbol{y} - v$.
  \STATE Let $\bbeta = u + \Lambda X^T\Sigma^{-1/2} w$.
\end{algorithmic}
\end{algorithm}

Clustering the variance parameters means that we can account for outlier observations if one of the variances $\sigma_{1}^{*},\ldots,\sigma_{K_{n}}^{*}$ is very large. Outlier observations
belonging to that specific cluster will have much higher variance than the others and can therefore take more extreme values. In this
way, the model can both simultaneously account for heteroskedasticity
and outliers which are characterised by clusters with extreme variances.

\subsubsection{Heavy Tails: Student-t extension} \label{sec:t-model}

In many real-world datasets, the response variable $\mathbf{y}$ may contain some larger than expected values which are commonly referred to as \textit{outliers}. A common approach to model such data is to replace the normal distribution assumption for the conditional distribution of $y_i$ given $\pmb{x}_i$ with a $\text{Student-t}$ distribution. Compared to a normal distribution, the Student-t distribution has heavier tails and therefore permits outlier observations with a higher probability than under the normal model. We can account for outliers in our Dirichlet process variable selection model \eqref{HSVS} by replacing the conditional distribution of $y_i$ with
$$
y_{i}|\boldsymbol{x}_{i},\boldsymbol{\beta},\boldsymbol{\sigma^{2}} \sim \text{Student-t}(\boldsymbol{x}_{i}^{\top}\boldsymbol{\beta},\sigma_{i}^2,\nu) \quad i=1,\ldots,n,
$$
for some degrees of freedom $\nu$. The parameter $\nu$ regulates the thickness of the tails of the distribution. A small value for $\nu$ leads to thicker tails with the expectation that large or extreme values for $y_i$ will be observed. As $\nu \rightarrow \infty$, the $\text{Student-t}$ distribution recovers the normal distribution.

A Gibbs sampler to perform posterior inference for the $\text{Student-t}$ model can be derived from Algorithm \ref{algo1} by including an additional data augmentation step. The required conditional distributions are derived by representing the Student-t distribution as a normal distribution with an inverse gamma variance. The remainder of this section describes these procedures. 

We assume our response variables are distributed as $Y\sim \text{Student-t}(\mu,\sigma^2,\nu)$ with density,
\begin{equation}
    f(y|\mu,\sigma^2,\nu)=\frac{\Gamma(\frac{\nu+1}{2})}{\Gamma(\frac{\nu}{2})\sqrt{\pi\nu\sigma^2}}\left(1+\frac{(y-\mu)^{2}}{\nu\sigma^2}\right)^{-\frac{\nu+1}{2}}
\end{equation}
with mean $\mathbb{E}(Y)=\mu$ and variance $\text{Var}(Y)=\sigma^2\frac{\nu}{\nu-2}$. A well-known reparameterization of the model follows that if, 
\begin{equation*}
    Y|G  \sim \mathcal{N}(\mu,G^{-1}) \ \ \mbox{and} \ \
     G  \sim \text{Gamma}(\nu/2,\nu \sigma^2/2),
\end{equation*}
then the marginal of $Y$ (when integrating out $G$) is  $Y\sim \text{Student-t}(\mu,\sigma^2,\nu)$. 

\begin{algorithm}
    \caption{DP variable selection Gibbs sampler, Student-t extension}
    \label{algo1_text}
\begin{algorithmic}
  \FOR{$t$ in 1: number of iterations }
  \FOR{i in 1:n }
        \STATE Sample classification vector $c_i$ with probability \eqref{c2}.
    %     \STATE Let $K^{-}$ be the number of distinct $c_{j}$ for $j\neq i$,  labeled $\{1,\ldots,K^{-}\}$. Draw new value $c_{i}$ from
    %     \begin{align*}
    %         \mathbb{P}[ c_{i} & =c\,|\,c_{-i},G_{1:n},\sigma^{2*}_{1:K_{n}}]\propto \\
    %   &  \propto \left\{\begin{array}{ll}
    %             n_{-i,c}\text{Ga}(G_{i};\nu/2,\nu \sigma_{c}^{2 *}/2) \hspace{0.2cm}\text{for }1\leq c \leq K^{-} \\[0.4cm]
    %             \alpha g(G_{i};\nu,b_{1},b_{2})\hspace{0.2cm}\text{for  }c=K^{-}+1,
    %         \end{array}\right.
    %     \end{align*} 
    %     where $n_{-i,c}$ is the number of $c_{j}=c$ for $j\neq i$.
        \ENDFOR
        
   \FOR{k in 1:$K_{n}$}
      \STATE  Sample $\sigma_{k}^{2*}$
       \begin{equation*}
       \sigma_{k}^{2*}\sim\mathrm{Gamma}\left(\frac{\nu}{2}n_{k}+b_{1},\frac{\nu}{2}\sum_{i:C_{k}}G_{i}+b_{2}\right),\;\label{eq:variance_sample2}
        \end{equation*}
        where $C_k = \{ G_i \ | \ c_i = k\}$.
   \ENDFOR

 \IF{Spike-and-Slab} \STATE Sample $\bbeta$, $\tau_j$, $\eta_j$ and $\omega$ using Algorithm \ref{alg:spike}, replacing $\Sigma=\mathrm{diag}(\sigma_{1:n}^{2})$  with $\Sigma =\mathrm{diag}(G_{1:n})^{-1}$. \ELSIF{Horseshoe} \STATE Sample $\bbeta$, $\lambda_j$ and $\tau$ using Algorithm \ref{alg:horseshoe}, replacing $\Sigma=\mathrm{diag}(\sigma_{1:n}^{2})$  with $\Sigma =\mathrm{diag}(G_{1:n})^{-1}$. \ENDIF
 
    \FOR{i in 1:$n$}
      \STATE  Sample $G_{i}$
       \begin{equation*}
       G_{i} \sim \text{Gamma}\left(\frac{\nu+1}{2},\frac{1}{2}[(Y_{i}-X_{i}^{T}\beta)^{2}+ \nu \sigma_{i}^{2}]\right).\;\label{eq:variance_sample3}
        \end{equation*}
   \ENDFOR

\STATE Sample $\alpha$, by sampling the latent variables
\begin{align*}
        \psi |\alpha,K_{n} &\sim \text{Beta}(\alpha+1,n) \ \ \mbox{then} \\
        a |\psi,K_n &\sim \text{Bernoulli}\left(\frac{w2}{w1+w2}\right) \\
        \alpha |a, \psi,K_{n} &\sim \text{Gamma}(d_{1}+K_{n} + a,d_{2}-\log \psi)
\end{align*}
where $w_1 = d_1 + K_n + 1$ and $w_2= n(d_2-\log\psi)$.
\ENDFOR
\end{algorithmic}
\end{algorithm}

The extension of the DPSS and DPHS models to a Student-t model is as follows
\begin{align}
    \label{eq:student-t-model}
    y_{i}|\boldsymbol{x}_{i},\boldsymbol{\beta},\sigma_{i}^{2} & \sim \text{Student-t}(\boldsymbol{x}_{i}^{\top}\boldsymbol{\beta},\sigma_{i}^2,\nu) \nonumber \\
    \boldsymbol{\beta} & \sim \pi_{\mathrm{SS/HS}} \nonumber \\
    \sigma_{i}^{2}|P & \sim P \\
    P & \sim \text{DP}(\alpha,\text{Gamma}(b_{1},b_{2})). \nonumber
\end{align}
Using the reparameterized Student-t model, and integrating out $P$ from \eqref{eq:student-t-model}, the joint posterior distribution $ \pi(\beta,G_{1:n},\sigma_{1:n}^{2},\alpha |Y_{1:n},X_{1:n})$ is proportional to, 
%To make inference on the model, we do data augmentation, introducing latent variables $G_{1},\ldots,G_{n}$. The augmented model is as follows,
% \begin{align*}
%     Y_{i}|X_{i},\beta,G_{i} & \sim N(X_{i}^{T}\beta,\frac{1}{G_{i
%     }}) \\
%     \beta & \sim \pi_{SS/HS}(\beta) \\
%     G_{i}|\sigma_{i}^{2} & \sim \text{Ga}(\frac{\nu}{2},\nu \frac{\sigma_{i}^{2}}{2}) \\
%     \sigma_{i}^{2}|P & \sim P \\
%     P & \sim \text{DP}(\alpha,\text{Ga}(b_{1},b_{2})).
% \end{align*}
%
% \begin{figure*}[ht]
% \includegraphics[scale=0.51]{fig/errors_p20_1_.png} \includegraphics[scale=0.51]{fig/errors_p20_3.1__.png}
% \caption{Estimation error $\|\hat{\beta}-\beta_0\|_2/\|\beta_0\|_2$ as a function of $n$ for fixed $p=20$ with homoskedastic (left) and heteroskedastic errors (right). \label{fig:err_p20}}
% \end{figure*}
\begin{align*}
  \prod_{i=1}^{n} & \mathcal{N}(y_{i}; \boldsymbol{x}_{i}^{T}\boldsymbol{\beta},G_{i
    }^{-1})\text{Gamma}(G_{i};\nu/2,\nu\sigma_{i}^{2}/2) \pi(\sigma_{1:n}|\alpha)\pi(\alpha)\pi_{SS/HS}(\beta),
\end{align*}
where as before, $\pi(\sigma_{1:n}|\alpha)$ denotes the marginal likelihood of $\sigma_{1:n}$ when $P$ is integrated out, which can be represented using the Chinese restaurant process. The proposed Gibbs sampler is specified in Algorithm \ref{algo1_text}, where at each iteration, we first resample a class vector $c_{1:n}$ assigning each $\sigma_{i}^{2}$ and $G_{i}$ to a block in the partition, and then resample the distinct values $\sigma_{1:k}^{2*}$ followed by each latent variable $G_{i}$. In the heavy tailed case, the observation is related to a variance through its latent variable and hence the posterior probability of the classification vector is
\begin{align}
    \mathbb{P}[ c_{i} & =c\,|\,c_{-i},G_{1:n},\sigma^{2*}_{1:K_{n}}]\propto \left\{\begin{array}{ll}
        n_{-i,c}\text{Gamma}(G_{i};\nu/2,\nu \sigma_{c}^{2 *}/2) \hspace{0.2cm}\text{for }1\leq c \leq K^{-} \\[0.4cm]
        \alpha g(G_{i};\nu,b_{1},b_{2})\hspace{0.2cm}\text{for  }c=K^{-}+1,
    \end{array}\right. \label{c2}
\end{align} 
where $K^{-}$ be the number of distinct $c_{j}$ for $j\neq i$, labeled $\{1,\ldots,K^{-}\}$, and $n_{-i,c}$ is the number of $c_{j}=c$ for $j\neq i$. Also, the likelihood of observation $i$ being assigned a new variance is modified to
\begin{align*}
 g(G_{i}; \nu,b_{1},b_{2}) & :=  \int \text{Gamma}(G_{i};\nu/2,\nu \sigma^{2}/2) \mathrm{Gamma}(\sigma^{2};b_{1},b_{2})d\sigma^{2}\\
 %& =  \int \frac{(\nu \sigma^2 /2)^{\nu/2}}{\Gamma(\nu/2)} G_{i}^{\frac{\nu}{2}-1} \exp(-G_{i}\frac{\nu\sigma^{2}}{2}) \cdot \\
 %& \hspace{2cm} \cdot \frac{b_{2}^{b_{1}}}{\Gamma(b_{1})} \sigma^{2(b_{1}-1)}\exp(-\sigma^{2} b_{2})d \sigma^{2}     \\
 %& =  \frac{(\nu  /2)^{\nu/2}}{\Gamma(\nu/2)} G_{i}^{\frac{\nu}{2}-1} \frac{b_{2}^{b_{1}}}{\Gamma(b_{1})} \int  \sigma^{2(b_{1}+\frac{\nu}{2}-1)}\cdot \\
 %& \hspace{2cm} \cdot \exp(-\sigma^{2} (b_{2}+ \frac{\nu G_{i}}{2})d \sigma^{2}  \\
 & =  \frac{(\nu  /2)^{\nu/2}}{\Gamma(\nu/2)} G_{i}^{\frac{\nu}{2}-1} \frac{b_{2}^{b_{1}}}{\Gamma(b_{1})} \frac{\Gamma(b_{1}+\frac{\nu}{2}) }{( b_{2}+ \nu G_{i}/2 )^{b_{1}+\frac{\nu}{2}}  }.
\end{align*}
Finally, regression coefficients $\bbeta$ are sampled using Algorithms \ref{alg:spike} and \ref{alg:horseshoe} with small modifications to include the augmented variable $G_i$.

\section{Experiments} \label{sec4}

In this section, we compare the DPSS and DPHS models against the standard spike-and-slab and horseshoe models of \citet{Ishwaran2005} and \citet{carvalho2010horseshoe}. We begin by studying their predictive performance, their ability to model heterogeneous data and their variable selection capabilities over a range of scenarios utilizing synthetic data, then we test the models' performance on a real-world dataset. %Firstly to assess the likelihood of predictions when modelling various open-source datasets and secondly 
Specifically, on the reconstruction of a network of genomic data through variable selection on a linear model. Code for synthetic and real experiments can be found in \url{https://github.com/albcab/RobustVariableSelection}.

For all experiments, each algorithm is run for $J=10,000$ iterations with a burn-in period of $J/2$. Convergence and mixing of the Markov chain is confirmed through visual diagnostics. Hyper-parameters for the models are set to $a_1=b_1=2.01$, $a_2=b_2=d_1=1$ and $d_2=1/2$, which leads to weakly informative priors. This claim is verified with simple cross validation for all the datasets. It is worth noting that one could place additional hyper-priors on these parameters and add extra steps to the MCMC algorithm. However, in doing so we would run the risk of rejecting steps potentially leading to longer mixing times. It is worth noting that in comparable studies \citep[e.g.][]{Ishwaran2005}, it is common to simply fix these parameters when assessing performance. Finally, both the dependent and independent variables are normalized to have zero mean for the testing of each model. 

%% FIG 1. Example of estimates and coefficient pattern
%\begin{figure}
%\includegraphics[width=1\columnwidth,height=4cm]{fig/beta_ex50.png}
%\caption{Example of coefficient pattern used in experiments, alongside an example of estimated coefficients for hSVS ($n=100$, $p=50$). \label{fig:sparsity_pattern}}
%\end{figure}

Performance is evaluated in the case of synthetic data over a range of regression scenarios varying the number of attributes $p=50,100,200, 500, 1000, 2000$ and data lengths $n=10,20,50,100,200$. The sparsity pattern will be kept similar across experiments according to the Briemann structure \citep{breiman2001statistical}. Specifically, we consider two scenarios for Gaussian linear regression: Scenario (1) - homoskedastic errors with $\epsilon\sim\mathcal{N}(0,1)$ as the single component and $\epsilon \sim \text{Student-t}(0, 1, 2)$ when testing the heavy tails extension; Scenario (2) - heteroskedastic errors with five components $\epsilon_{i}\sim \mathcal{N}(0,\sigma^2_i)$ or $\epsilon \sim \text{Student-t}(0, \sigma^2_i, 2)$ for $\sigma^2_i \in \{0.5, 1, 1.5, 2, 2.5\}$ distributed evenly among the observations as well as $1,2$ and $4$ outliers $\epsilon\sim\mathcal{N}(0,10)$ or $\epsilon \sim \text{Student-t}(0, 10, 2)$ in the case of $n=50,100,200$, respectively.  
% \[
% \epsilon_{i}\sim\begin{cases}
% \mathcal{N}(0,0.5) & \mathrm{if}\;1\le i\le n/5\\
% \mathcal{N}(0,1) & \mathrm{if}\;n/5<i\le2n/5\\
% \mathcal{N}(0,1.5) & \mathrm{if}\;2n/5<i\le 3n/5\\
% \mathcal{N}(0,2) & \mathrm{if}\;3n/5<i\le4n/5\\
% \mathcal{N}(0,2.5) & \mathrm{if}\;4n/5<i\le n,
% \end{cases}\;
% \]

In each scenario, and for each combination of attributes and data lengths, 100 synthetic datasets are created and parameters are estimated for each of these datasets. The results presented in the remainder of this section consider all 100 estimates for robustness.

% \begin{figure*}[ht]
%  \includegraphics[scale=0.51]{fig/errors_p200_1__.png} \includegraphics[scale=0.51]{fig/errors_p200_3.1_.png}
% \caption{Estimation error $\|\hat{\beta}-\beta_0\|_2/\|\beta_0\|_2$ as a function of $n$ for fixed $p=200$ with homoskedastic (left) and heteroskedastic errors (right). \label{fig:err_p200}}
% \end{figure*}

\subsection{Homoskedastic and Heteroskedastic Errors}
\label{sec:homo-hetero}

\begin{figure*}[ht]
 \includegraphics[width=1.\columnwidth,height=1.35\linewidth]{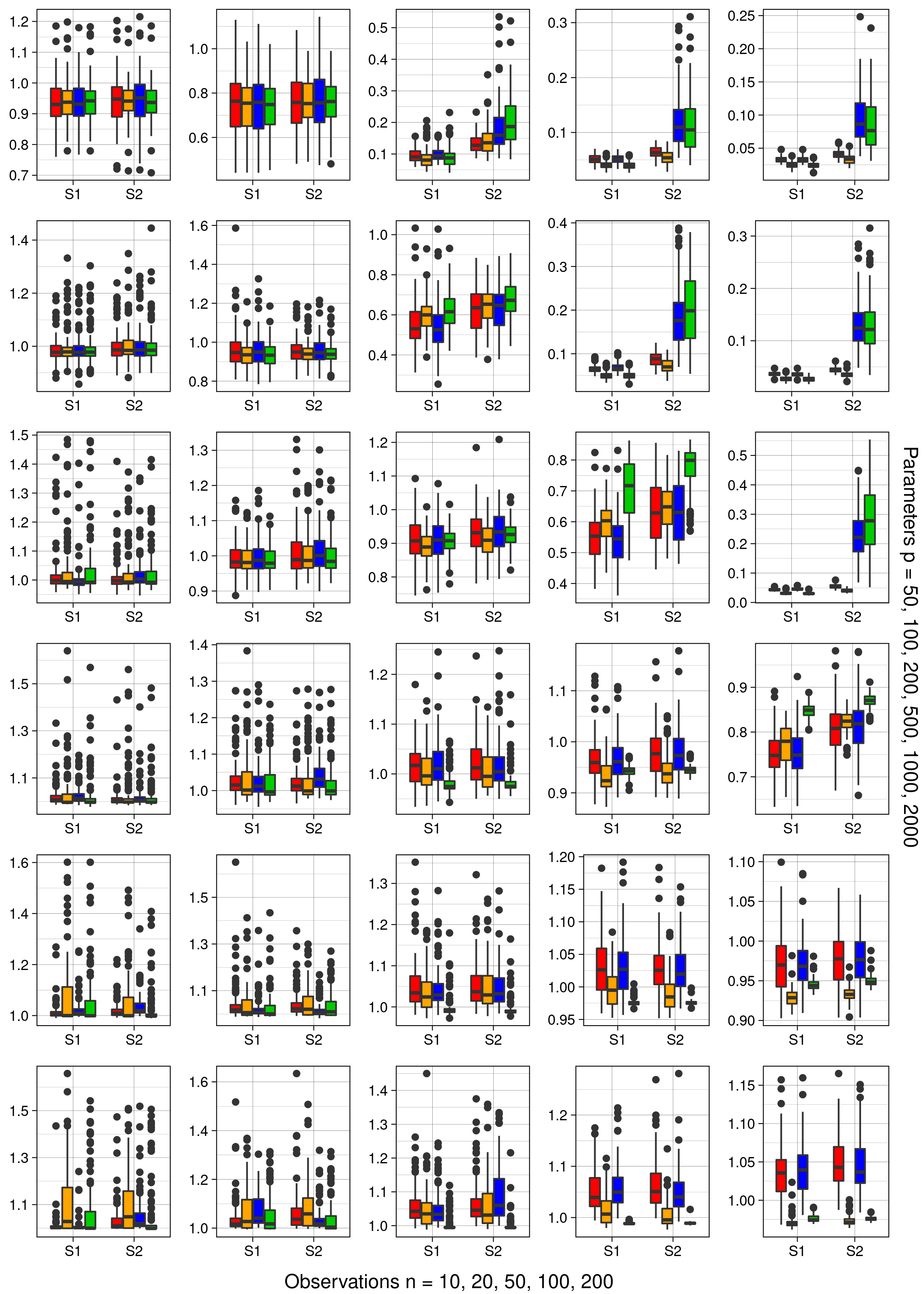}
\caption{Boxplots of estimation error $\|\hat{\beta}-\beta_0\|_2/\|\beta_0\|_2$ on \textit{Scenario 1} (S1) and \textit{Scenario 2} (S2) for algorithms \textcolor{red}{Dirichlet process horseshoe}, \textcolor{yellow}{Dirichlet process spike-and-slab}, \textcolor{blue}{Horseshoe}, and \textcolor{green}{Spike-and-slab}. \label{fig:error}}
\end{figure*}

\begin{figure*}[ht]
 \includegraphics[width=1.\columnwidth,height=1.35\linewidth]{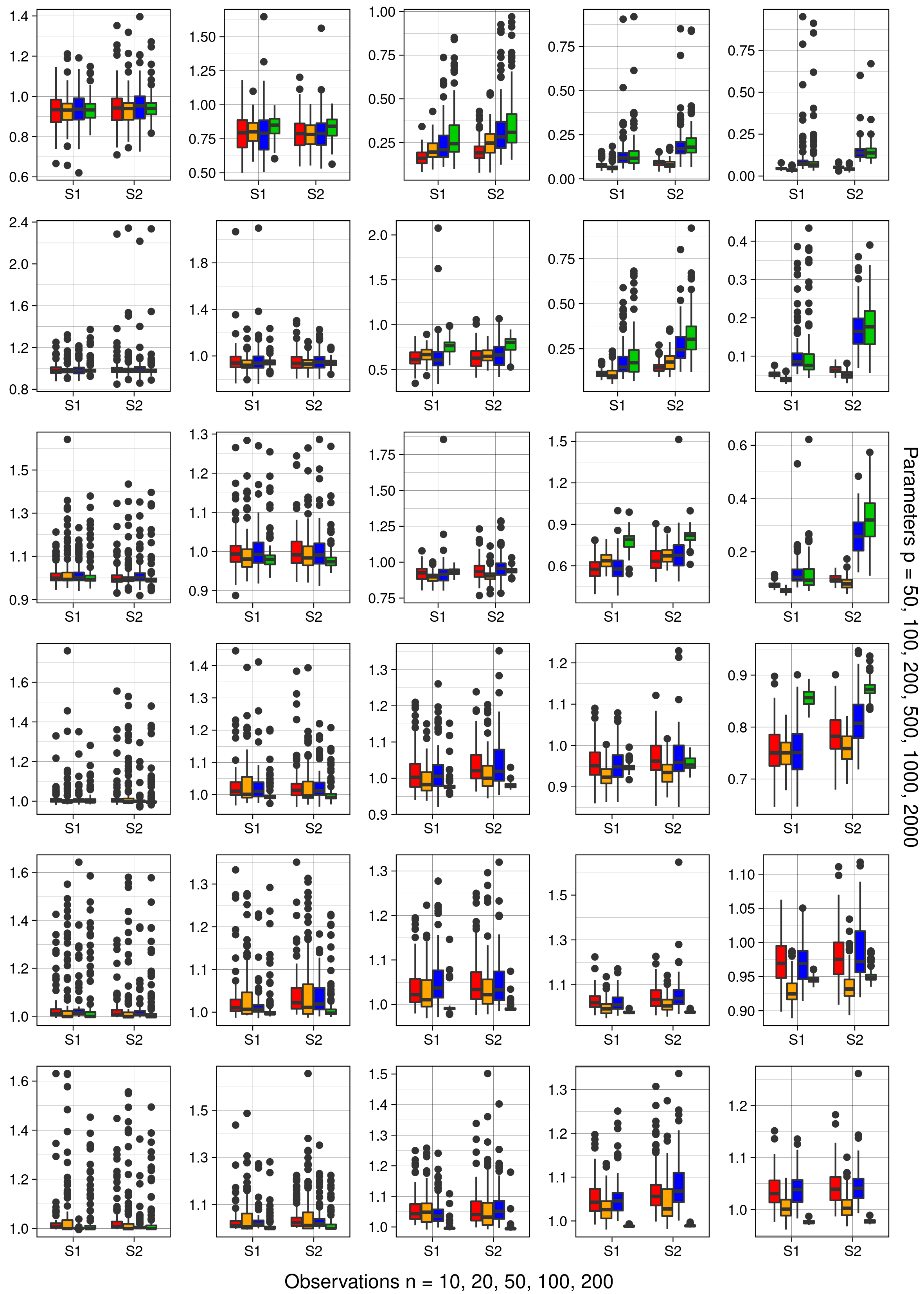}
\caption{Boxplots of estimation error $\|\hat{\beta}-\beta_0\|_2/\|\beta_0\|_2$ on \textit{Scenario 1} (S1) and \textit{Scenario 2} (S2) for algorithms \textcolor{red}{Dirichlet process horseshoe}, \textcolor{yellow}{Dirichlet process spike-and-slab}, \textcolor{blue}{Horseshoe}, and \textcolor{green}{Spike-and-slab} on their heavy tailed Student-t extension. \label{fig:error_tstud}}
\end{figure*}

We compare the Dirichlet process variable selection models against the standard spike-and-slab and horseshoe models in the presence of homoskedastic (\textit{Scenario 1}) and heteroskedastic (\textit{Scenario 2}) noise. Figure \ref{fig:error} presents boxplots of the posterior error $\|\hat{\beta}-\beta_0\|_2/\|\beta_0\|_2$ for every combination of number of observations $n$ and parameters $p$, with \textit{Scenario 1} plotted on the left and \textit{Scenario 2} given on the right, while Figure \ref{fig:error_tstud} presents the same error boxplots when testing the heavy tailed extension of the models. 

\textbf{Robustness} - it is clear from the upper right plots on Figure \ref{fig:error} that for \textit{Scenario 2} the robustness provided by the DPSS and DPHS models improves the predictive estimates of parameters $\bbeta$ when data is sufficient to make the model overdetermined, i.e. when $n$ is at least as large as $p$. As expected, under \textit{Scenario 1} the results are similar for the models allowing heterogeneous and homogeneous variances. Interesting behaviour is observed when the model is underdetermined at varying degrees. When the model is highly underdetermined $n \ll p$, models with the horseshoe priors provide more robust results in comparison to spike-and-slab priors. However, as $n$ approaches $p$ the spike-and-slab prior provides lower errors than the horseshoe (albeit at a smaller scale), until the model becomes overdetermined and the horseshoe prior models are preferred. Similar results are observed for the heavy tailed extension of the models. In contrast to the Gaussian models, when the models are at least determined $n \geq p$ the positive effect of robust variance modelling is less noticeable. 

% when the parameters are low-dimensional the robustness provided by the DPSS and DPHS models improves the predictive estimates when data is scarce. While in the high-dimensional setting, the Dirichlet process construction is able to extract additional information from the larger datasets. In a similar fashion, under \textit{Scenario 1}, the robust models provide better estimates when the dimension is low and utilizes new data better in the case of large $p$.

% \iffalse
% %% FIG 2. Standard dimensional single variance
% \begin{figure}[ht]
% \includegraphics[width=1\columnwidth,height=4cm]{fig/errors_p20_1.png}
% \includegraphics[width=1\columnwidth,height=4cm]{fig/errors_p200_1.png}
% \caption{Estimation error $\|\hat{\beta}-\beta_0\|_2/\|\beta_0\|_2$ for both SVS and hSVS as a function of $n$ for fixed $p=20$ (Top) and $p=200$ (Bottom. \label{fig:err_n_low}}
% \end{figure}
% \fi

% \iffalse
% %% FIG 5. Scaling of estimation error as function of $n$, fixed $p=20$
% \begin{figure}
% \includegraphics[width=1\columnwidth,height=4cm]{fig/errors_p20_3.png}
% \includegraphics[width=1\columnwidth,height=4cm]{fig/errors_p200_3.png}
% \caption{Normalised posterior error $\|\hat{\beta}-\beta_0\|_2/\|\beta_0\|_2$ as a function of $n$ for fixed $p=20$ (Top) and $p=200$ (Bottom). \label{fig:err_p}}
% \end{figure}
% \fi

\textbf{Coefficients} - looking at Figure \ref{fig:error} and \ref{fig:error_tstud} from top to bottom the effect of adding coefficients to the model is illustrated on the predictive capabilities of the models. The Dirichlet process model performs well in the presence of heterogeneity, particularly as the number of model coefficients is less than the number of observations. The robustness added by the heterogeneous variances benefits the horseshoe prior in the overdetermined case and the spike-and-slab prior in the underdetermined case.

%% FIG 7. Scaling of estimation error as function of $p$, fixed $n=100$ 
% \begin{figure}
% \includegraphics[width=1\columnwidth,height=4cm]{fig/errors_n100_3.1.png}
% \caption{Normalised posterior error $\|\hat{\beta}-\beta_0\|_2/\|\beta_0\|_2$ as a function of dimension $p$ for fixed $n=100$. \label{fig:err_n}}
% \end{figure}

\begin{figure}
\includegraphics[width=.7\columnwidth,height=6cm]{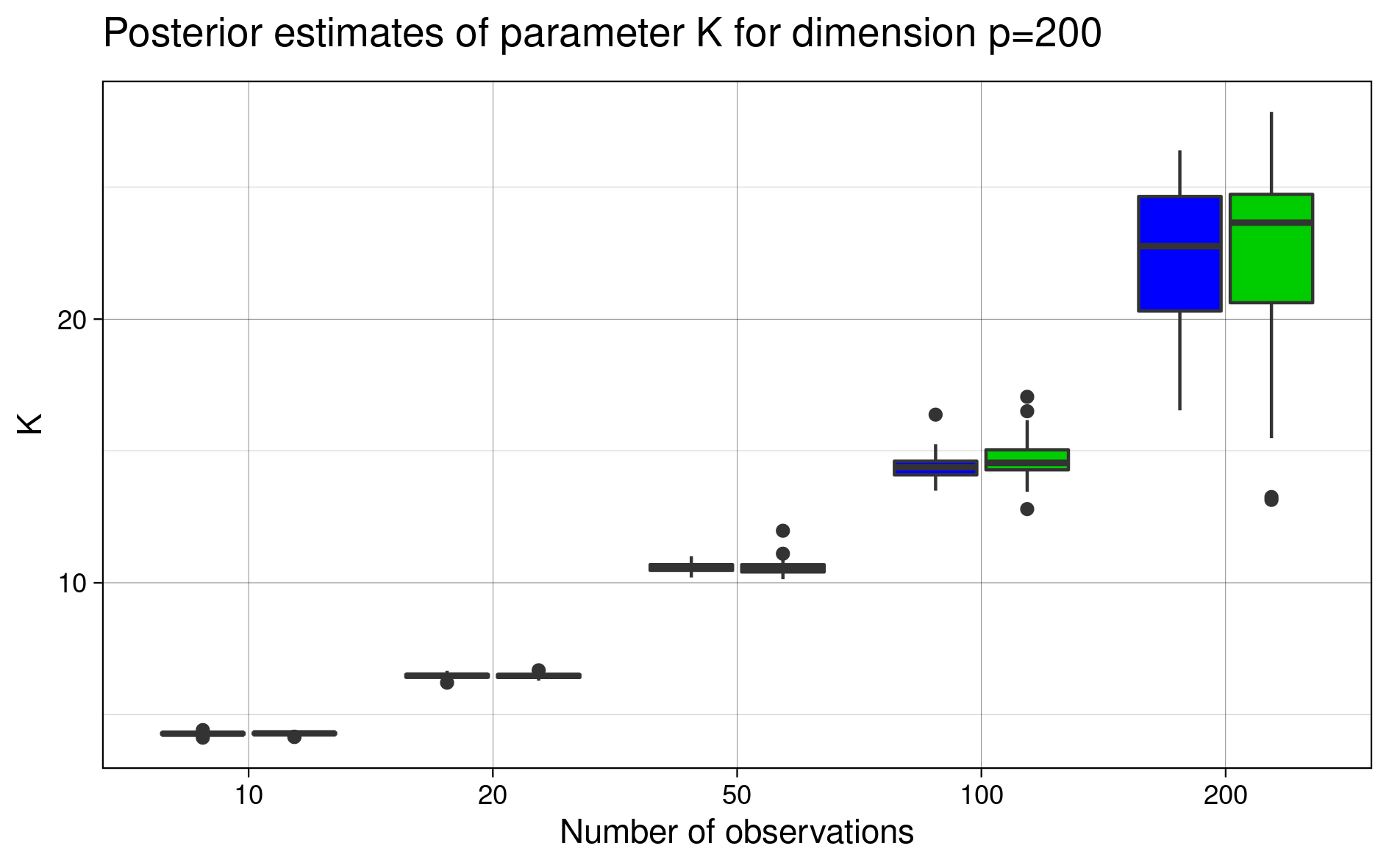}
\caption{Boxplots of the derived number of clusters $K$ from the posterior of $\sigma_{1},\ldots,\sigma_{n}$ for $p=200$ for algorithms \textcolor{blue}{Dirichlet process horseshoe}, \textcolor{green}{Dirichlet process spike-and-slab}, where the true number of clusters is $K=5, 5, 6, 7, 9$ for $n=10,20,50,100,200$.} \label{fig:clusterK}
\end{figure}

\begin{figure}
\includegraphics[width=.7\columnwidth,height=6cm]{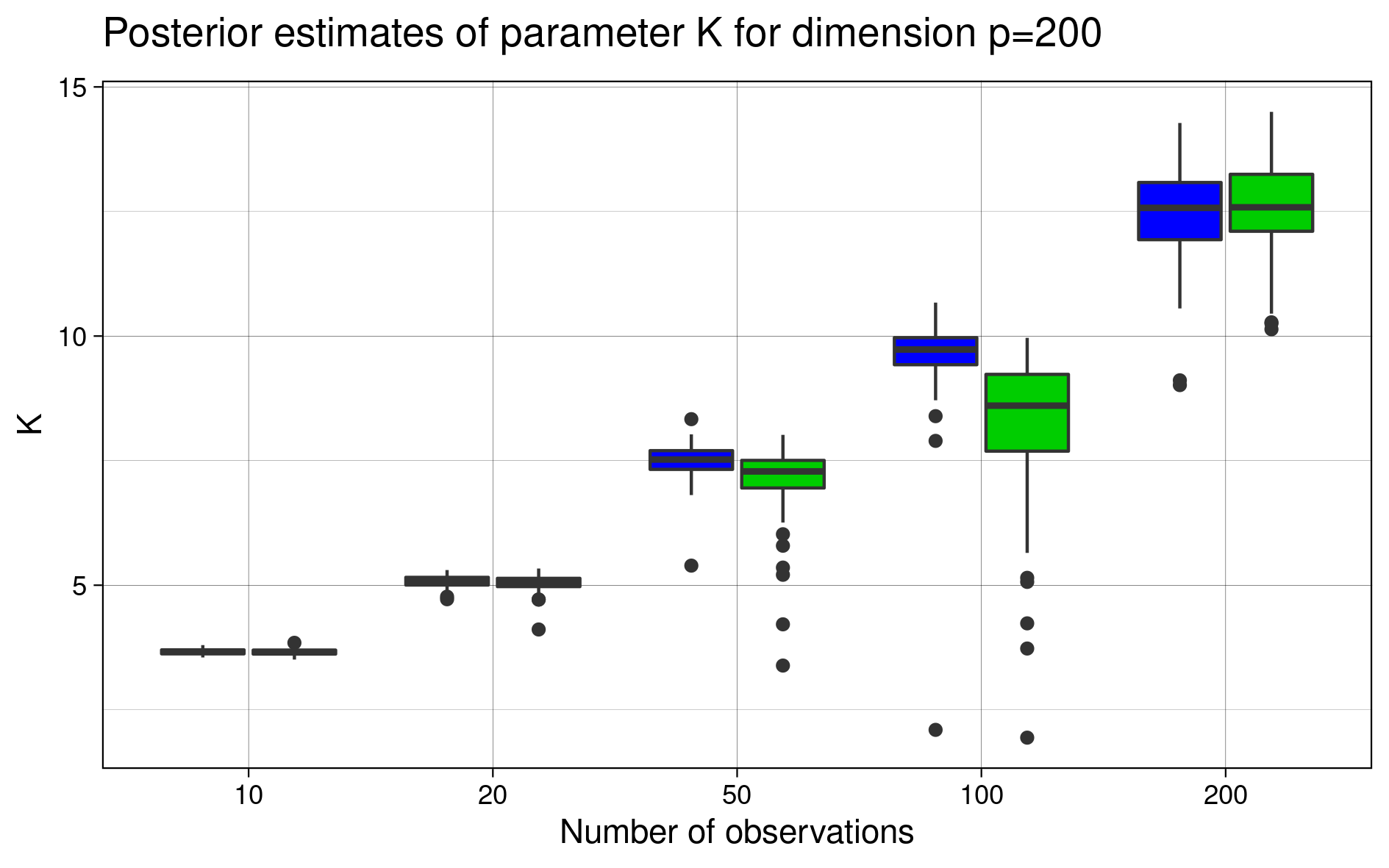}
\caption{Boxplots of the derived number of clusters $K$ from the posterior of $\sigma_{1},\ldots,\sigma_{n}$ for $p=200$ for algorithms \textcolor{blue}{Dirichlet process horseshoe}, \textcolor{green}{Dirichlet process spike-and-slab} on their heavy tailed Student-t extension, where the true number of clusters is $K=5, 5, 6, 7, 9$ for $n=10,20,50,100,200$.} \label{fig:clusterK_tstud}
\end{figure}

\begin{figure}
\includegraphics[width=0.49\columnwidth]{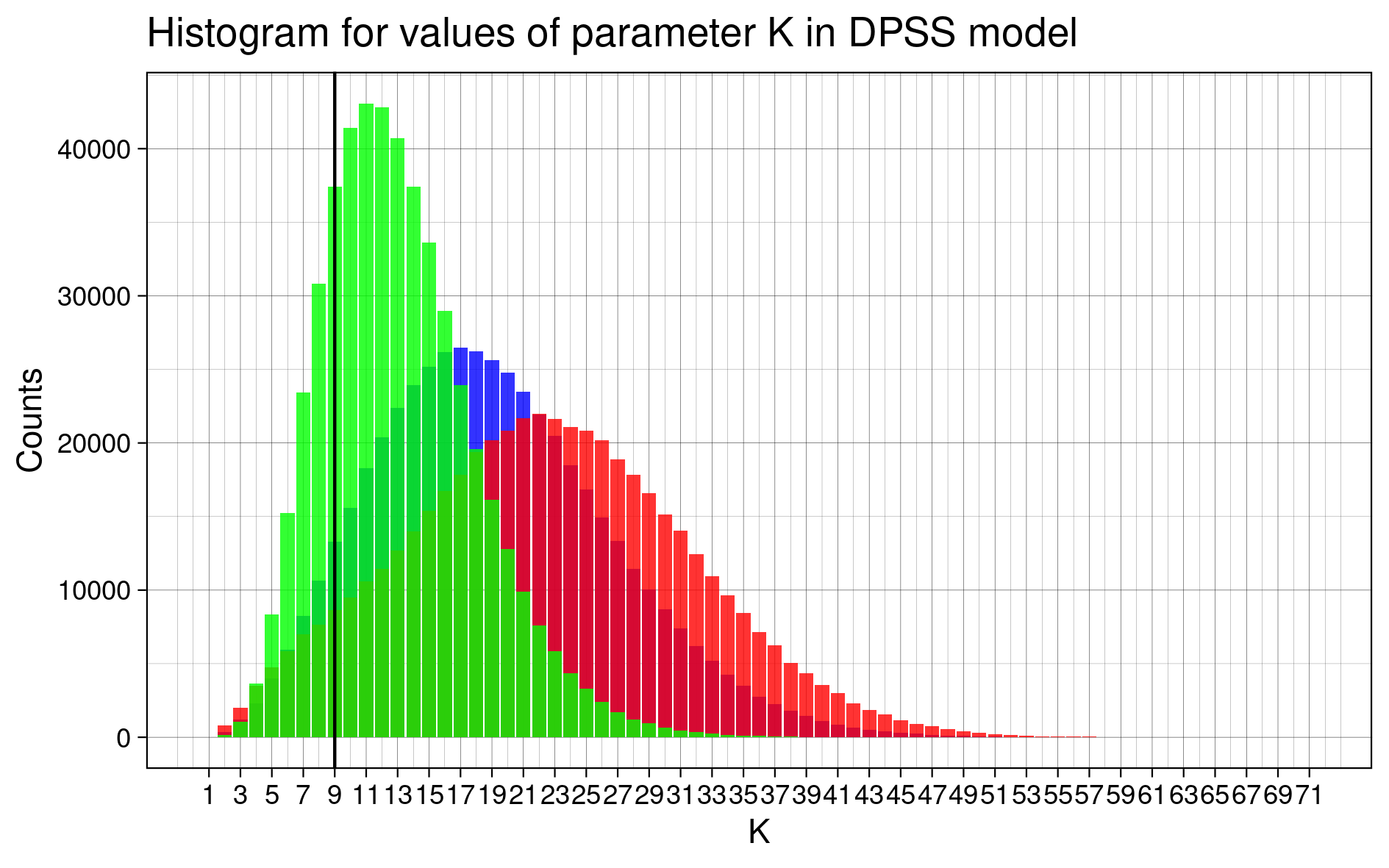}
\includegraphics[width=0.49\columnwidth]{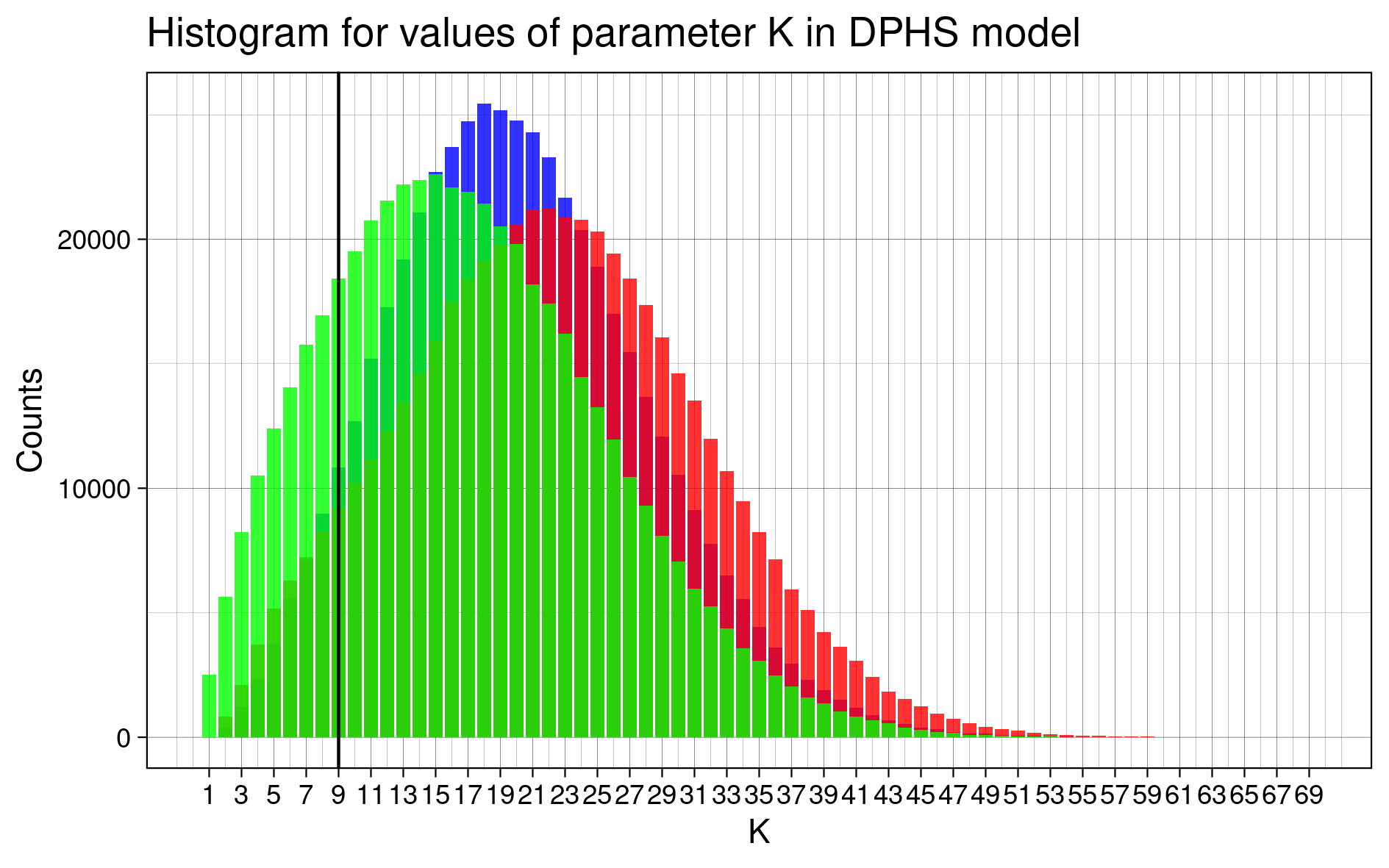}
\caption{Histogram of the derived number of clusters $K$ from the posterior of $\sigma_{1},\ldots,\sigma_{n}$ for $n=200$ and $p= $ \textcolor{blue}{50}, \textcolor{red}{200}, and \textcolor{green}{2000}, where the two number of clusters is $K=9$.} \label{fig:cluster_convergence}
\end{figure}

\begin{figure}
\includegraphics[width=0.49\columnwidth]{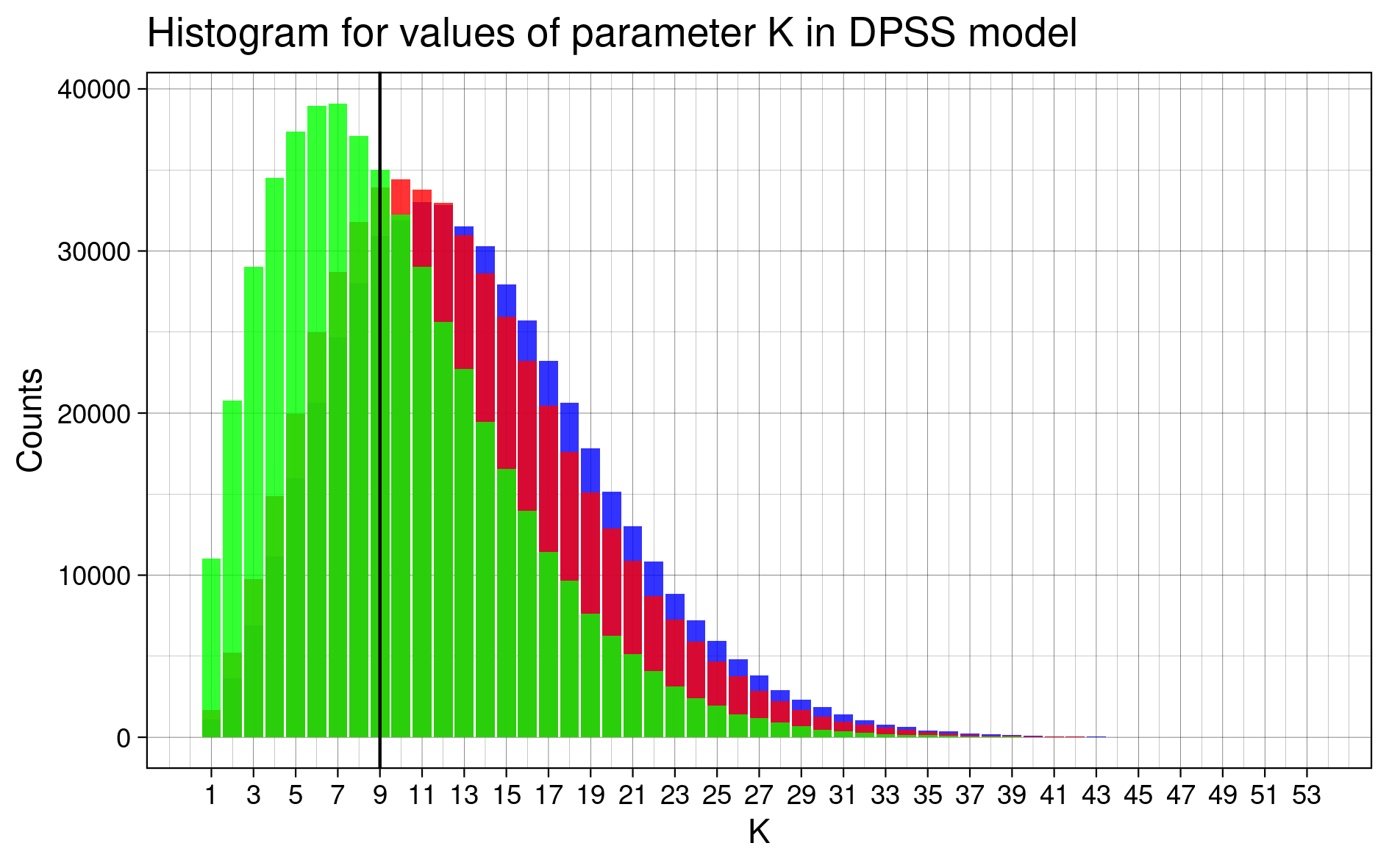}
\includegraphics[width=0.49\columnwidth]{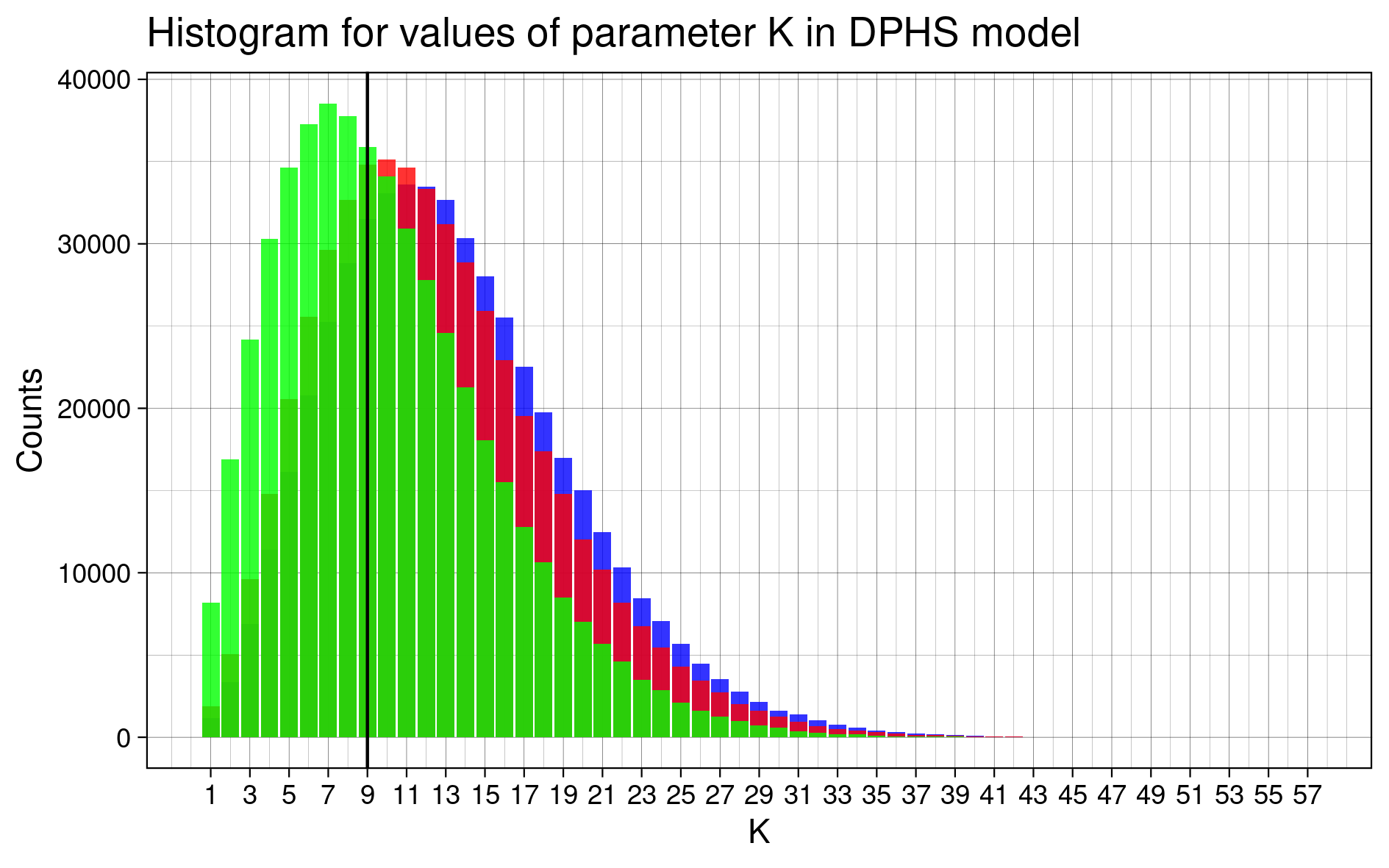}
\caption{Histogram of the derived number of clusters $K$ from the posterior of $\sigma_{1},\ldots,\sigma_{n}$ for $n=200$ and $p= $ \textcolor{blue}{50}, \textcolor{red}{200}, and \textcolor{green}{2000} on their heavy tailed Student-t extension, where the two number of clusters is $K=9$.} \label{fig:cluster_convergence_tstud}
\end{figure}

\textbf{Clusters} - one of the key advantages of the nonparametric approach we adopt is the availability 
of posterior densities for the number of variance components $K$. The Dirichlet process' assumption of a number of clusters which scale logarithmically towards infinity with $n$ can be observed by looking at the boxplots for samples of $K$ for an increasing $n$ in Figures \ref{fig:clusterK} and \ref{fig:clusterK_tstud} for the heavy tailed extension. A potentially useful caveat of the heavy tailed extension of the robust models is that the number of clusters $K$ increases at a slower rate than the case of Gaussian errors. In the case where a finite number of clusters in the data is known, or the number of clusters $K$ should scale faster or slower than logarithmically, priors that generalize the Dirichlet process can be used, e.g. the Pitman-Yor process \citep{pitman1997two}. It is interesting to notice how, as the number of coefficients increases but the number of observations remain constant, the number of clusters $K$ retract themselves on their bias when the number of coefficients becomes much larger than the number of observations, as shown in Figures \ref{fig:cluster_convergence} and \ref{fig:cluster_convergence_tstud}. A comparison of these two Figures also shows how the heavy tailed extension of the robust model provides a more gradual increase on the number of clusters $K$.

%% FIG 8. Histogram over K for different $n-50,100,200$ with $p=100$ 

%% FIG 8. Histogram over K for different $n-50,100,200$ with $p=100$ 

\subsection{Support Recovery} \label{support}

In addition to estimating the regression coefficients, a further task of interest, especially in high-dimensional regression, is to select a subset of attributes which are deemed significant for the predictive model. In a frequentist context, this is usually achieved via forward selection with sequential F-tests, or with $\ell_{1}$ or $\ell_{0}$ shrinkage and/or selection, for instance through the use of the AIC or BIC criteria.

One advantage of the Bayesian approach is that we can sample from the joint posterior of the coefficients, thus we can construct credible intervals with relative ease. In this section, we compare two methods for estimating the support of the regression model:

\begin{itemize}[noitemsep]
\item The first approach is to simply look at the posterior of the inclusion parameter $\eta_{j}$ for selecting either the spike, or slab. Specifically, if the mode of the posterior $\hat{\eta}_{j}$ is above level $1-\zeta$ we add the index $j$ to the estimated support set $\hat{\mathcal{M}}$. This method works only in the DPSS and SS models.
\item The second approach is to look at the empirical posterior credible interval of $\hat{\beta}_{i}$ at the percentiles $\zeta/2\times100$ and $(1-\zeta/2)\times100$. If the constructed interval excludes zero, then we add the index to the estimated support set $\hat{\mathcal{M}}$. This method is used for the DPHS and HS models.
\end{itemize}

We note that the second approach is somewhat similar to the \textit{z-cut} method discussed in \citet{Ishwaran2005}, however, they construct a set $\hat{\mathcal{M}}:=\{i\;|\;\forall i\;s.t.\;|\bar{\beta}_{i}|\\ \ge z_{(1-\zeta/2)}\}$
where $\bar{\beta}_{i}$ represents the posterior mean of the regression coefficients.

To summarise the performance of the two approaches, we take the number of true and false positive ($\mathrm{TP}$, $\mathrm{FP}$) coefficients included in the model and consider how this changes as a function of $n$. Clearly, these values will change as a function of $\zeta$, and in practice this may be tuned to favour either TP or FP results. For simplicity, we report results with $\zeta=0.05$. Table \ref{table:support} summarises the performance of the posterior inclusion thresholding method for different values of $n$ in the heteroskedastic scenario (\textit{Scenario 2}) and the standard Gaussian formulation of the models, similar results where observed on their heavy tailed extension. In the case of an overdetermined model, both the standard and robust implementations learn correctly the underlying data generating process. However, when the model is underdetermined the standard spike-and-slab tends to be more conservative in the inclusion of variables while the robust method favour a better estimate TP with the drawback of more FP. Similar results can be observed in the case of the horseshoe prior. The support recovery, or model selection, of the robust models will be detailed further when applied to real world data, where the benefits of each prior are illustrated.

\begin{table*}[t] 
\centering
\begin{tabular}{llrlrlrlrl} 
  \hline
Model & n & 20 &  & 50 &  & 100 &  & 200 &  \\ 
  \hline
SS & TP & 5 & [0.95,8] & 14 & [12,14] & 14 & [14,14] & 14 & [14,14] \\ 
   & FP & 3 & [0,7] & 0 & [0,1] & 0 & [0,0] & 0 & [0,0] \\ 
  DPSS & TP & 5 & [2,8] & 14 & [12,14] & 14 & [14,14] & 14 & [14,14] \\ 
   & FP & 3 & [0.95,8] & 0 & [0,1] & 0 & [0,1] & 0 & [0,0] \\ 
   \hline
  SS & TP & 4 & [1,10] & 17 & [1,25] & 28 & [28,28] & 28 & [28,28] \\ 
   & FP & 5 & [0,13.15] & \textbf{18} & [0,43.3] & 0 & [0,1] & 0 & [0,1] \\ 
  DPSS & TP & 4 & [1,9.05] & \textbf{20} & [14,26] & 28 & [28,28] & 28 & [28,28] \\ 
   & FP & 4 & [0,14] & 24 & [6.95,46.05] & 0 & [0,1] & 0 & [0,1] \\ 
   \hline
  SS & TP & \textbf{4} & [0,18.15] & 25 & [0,51.05] & 31 & [1,56] & 56 & [56,56] \\ 
   & FP & 6 & [0,36.55] & \textbf{53} & [0,124.2] & \textbf{62} & [0,144] & 0 & [0,1] \\ 
  DPSS & TP & 3 & [0,12.1] & \textbf{42} & [16.95,56] & \textbf{54} & [47.9,56] & 56 & [56,56] \\ 
   & FP & 6 & [0,22] & 99 & [25.9,142.05] & 133 & [95.8,144] & 0 & [0,1] \\ 
   \hline
\end{tabular}
\caption{Tabulated results of true positive (TP) and false positive (FP) results with 95\% credible intervals for the inclusion of regression coefficients for $p=50, 100, 200$ with nonzero coefficients TP $=14, 28, 56$.} \label{table:support}
\end{table*}

\subsection{Reconstruction of transcription regulatory networks}

We consider the problem of reconstructing genetic regulatory networks from gene expression data \citep{marbach2010revealing}. This problem can be modelled as a directed network, where each node corresponds to a different gene and each connection represents a directed interaction between two genes at the transcription level. The sparse spike-and-slab and horseshoe priors provide an attractive approach to reconstructing these networks using support recovery methods.

The models are tested on data from the challenge posed at the \textit{The Dialogue for Reverse Engineering Assessments and Methods} (DREAM) 2009 conference, specifically the multifactorial subchallenge\footnote{www.synapse.org/\#!Synapse:syn3049712/wiki/74628}. The challenge consists of reverse engineering five independent networks using 100 steady-state measurements for each network with 100 genes. The levels of expression of all the genes are measured under different perturbed conditions. Each gene $X_i$ for $i=1,...,100$ is treated independently and modelled through a linear relationship with the rest of the genes in the network $\boldsymbol{X}$, i.e. $X_{i} = \boldsymbol{X}^T\boldsymbol{\beta}_i + \epsilon$. Every $\boldsymbol{\beta}_i$ has an independent spike-and-slab or horseshoe prior and $\epsilon$ is a normal random variable with mean zero and homogeneous variances in the case of SS and HS, or heterogeneous variances, in the case of DPSS and DPHS. For each gene, the support is recovered using the method described in Section \ref{support} for the case of a spike-and-slab prior with $\boldsymbol{\beta}_i$. In the case of a horseshoe prior, we follow \citet{steinke2007experimental} and approximate the posterior probability of a connection from gene $j$ to gene $i$ by the probability
of the event $|(\boldsymbol{\beta}_i)_j| > 0.1$ under the posterior for $\boldsymbol{\beta}_i$.

The performance of the different approaches is evaluated using the mean of the logarithmic loss of the probabilities of connections $p_{ij}$ from gene $j$ to each gene $i$, defined as
$$\frac{1}{100}\sum_{i=1}^{100} \left\{ - \frac{1}{99} \sum_{j \neq i} [y_{ij} \log \, p_{ij} + (1 - y_{ij}) \log \, (1 - p_{ij})] \right\},$$
where $y_{ij} = 1$ if there is a directed connection from $j$ to $i$ and $y_{ij} = 0$ otherwise. Table \ref{loss} presents these results for each tested model. The improvement in prediction given by heterogeneous variances in the model is substantial. This supports the results from Section \ref{support} on synthetic data, illustrating the improvement offered by robust models to correctly classify variables included in the model, specifically in their capability of reducing type II error in the predictions while correctly identifying the significant coefficients in the linear relationship. It is interesting to note that, while the DPSS model provided better support recovery results in the synthetic data example, the DPHS provides better results in the real data example of gene network reconstruction.

\begin{table}[h]
\centering
\begin{tabular}{lrr|rr}
  \hline
  & SS & DPSS & HS & DPHS \\
  \hline
N1 & 0.1143 & \textbf{0.0948} & 0.0920 & \textbf{0.0854} \\ 
  N2 & 0.1498 & \textbf{0.1380} & 0.1248 & \textbf{0.1173} \\ 
  N3 & 0.1610 & \textbf{0.1158} & 0.1196 & \textbf{0.0900} \\ 
  N4 & 0.1588 & \textbf{0.1118} & 0.1124 & \textbf{0.0970} \\ 
  N5 & 0.1314 & \textbf{0.1095} & 0.1044 & \textbf{0.0933} \\ 
 \hline
\end{tabular}
\caption{Logarithmic-loss errors for the five DREAM Gene Network detection datasets (N1-N5).}
\label{loss}
\end{table}

% \begin{table}[h]
% \centering
% \begin{tabular}{lrr|rr}
%   \hline
%   & SS & DPSS & HS & DPHS \\
%   \hline
% N1 & \textbf{0.1549} & 0.1620 & \textbf{0.1545} & 0.1592 \\
%   N2 & 0.2013 & \textbf{0.1822} & 0.1770 & \textbf{0.1697} \\
%   N3 & \textbf{0.2161} & 0.2215 & \textbf{0.1998} & 0.2112 \\
%   N4 & 0.2395 & \textbf{0.1709} & 0.2143 & \textbf{0.1701} \\
%   N5 & \textbf{0.1857} & 0.2044 & \textbf{0.1959} & 0.1984 \\
%  \hline
% \end{tabular}
% \caption{Logarithmic-loss errors for the five DREAM Gene Network detection datasets (N1-N5).}
% \label{losst}
% \end{table}

\section{Conclusion}
This paper outlines a nonparametric extension of popular Bayesian variable selection models to account for heterogeneity, outliers and clustering effects. We also provide an extension to the model to allow for heavy-tailed data  distributions. Fitting our Dirichlet process variable selection models is computationally efficient using a Gibbs sampling construction. The results presented for both synthetic and real data examples show a robust improvement in both predictive accuracy on test data and improved efficiency in identifying key model attributes.

This work could be further extended to encompass other variable selection models, including extensions of existing models, such as the regularized horseshoe model \citep{piironen2017sparsity}. It would also be interesting to explore nonparametric priors that allow for more flexible assumptions on the expected clustering effect, such as the Pitman-Yor process \citep{pitman1997two}.

%%%%%%%%%%%%%%%%%%%%%%%%%%%%%%%%%%%%%%%%%%%%%%
%% Supplementary Material, if any, should   %%
%% be provided in {supplement} environment  %%
%% with title and short description.        %%
%%%%%%%%%%%%%%%%%%%%%%%%%%%%%%%%%%%%%%%%%%%%%%
% \begin{supplement}
% \stitle{Title of Supplement A}
% \sdescription{Short description of Supplement A.}
% \end{supplement}
% \begin{supplement}
% \stitle{Title of Supplement B}
% \sdescription{Short description of Supplement B.}
% \end{supplement}

\bibliographystyle{ba}
\bibliography{refs}

\begin{acks}[Acknowledgments]
CN gratefully acknowledges the support of EPSRC grants EP/V022636/1, EP/S00159X/1 and EP/R01860X/1. 
\end{acks}

\end{document}